# THE EVOLUTION OF COMMUNICATION SYSTEMS

Loet Leydesdorff

Department of Science Dynamics, Nieuwe Achtergracht 166, 1018 WV Amsterdam,

The Netherlands

[(return)](#)

*Abstract*

One can study communications by using Shannon's (1948) mathematical theory of communication. In social communications, however, the channels are not "fixed", but themselves subject to change. Communication systems change by communicating information to related communication systems; co-variation among systems if repeated over time, can lead to co-evolution. Conditions for stabilization of higher-order systems are specifiable: segmentation, stratification, differentiation, reflection, and self-organization can be distinguished in terms of developmental stages of increasingly complex networks. In addition to natural and cultural evolution, a condition for the artificial evolution of communication systems can be specified.

*Introduction*

Evolution theory assumed traditionally that the "natural" environment selects. From this perspective the natural environment is an external given for the evolving system, which itself can exhibit only variation. If selection, however, feeds information back into the evolving system, the environment can no longer be conceptualized as a given, but it must be considered as another communication system that exhibits variation. The system/environment relation is consequently a relation between communication systems. The communication systems inform each other by communicating.

In general, communication systems can communicate information only with other communication systems. Communication systems communicate through "mutual information" or co-variation. When a pattern of co-variation among them is maintained over time, systems may begin to co-evolve, i.e., mutually to shape one another. Co-evolution, and not evolution, is consequently the general concept for understanding dynamic developments. The concept of co-evolution enables us to understand, among other things, how new information can enter a system from its environment.

In traditional evolution theory, "natural" selection was supposed to lead to the survival of specific variants. In the case of co-evolution theory, the *stabilization* of specific co-evolutions adds a third mechanism to the previous pair of variation and selection. Among these three mechanisms (variation, selection, and stabilization) at least two cybernetics can be defined. While selection can occur at discrete moments in time, stabilization presumes the assessment of variation and selection over the time dimension. Stabilization is therefore a higher-dimensional problem. I shall show that the possibility of stabilization *and* self-organization can be considered as a consequence of the recursivity of the selective operation.

*Variation and Selection*

Let us first focus on the relation between "variation" and "selection." By operating, the systems vary their relations with systems in their environments, and thus inform each other. Each system can process this information internally for a self-referential update, if it is structurally able to do so (see below). Externally, i.e., by crossing the system/environment boundary, the information becomes part of a transmitting system, but this system transmits the information as a message. The nature of this message is specific for the transmitting system, and the originally sent information is packaged into this message. A transmitting system communicates messages as *its* information, but this information is different from the originally sent information. For example, the telephone transforms the spoken communication on the telephone line into a message in terms of electrical currents.

Thus, the communication systems are substantively different, and they can be delineated in terms of *what* they communicate. The various substances interact exclusively in terms of co-variations that function as windows of communication. The co-variation is part of two different variations in otherwise orthogonal dimensions. The dimensions stand analytically in orthogonal relations to one another, since the communicating systems are substantively different. This does not preclude that they can both be part of a supersystem, but I shall show below that this involves a third dimension.

In other words, each communication triggers a communication in the communication system(s) to which the sending system relates, since communication implies co-variation. Each communication system may co-vary with various other communication systems; each co-variation adds another degree of freedom to the communication within the respective systems. A co-variation, however, is part of the total variation of the co-varying systems. Through the co-variation the systems mutually inform each other. The mutual information or the co-variance, therefore, can be used as measures of the communication.[Note 1] A co-variance is always the complement of a remaining variance to a total variance. The remaining variance is itself a summation of co-variances that represent co-variations at other moments. In other words, the information content of communication systems is nothing but an expectation of communication in various dimensions.

When compared with a previous stage, the remaining variance is stable in relation to the co-variance in a specific instance. As noted, stability requires assessment over time as a separate dimension. The remaining variance at each *a posteriori* moment can be considered as auto-covariance in the time dimension. Auto-covariance indicates self-referentiality, and thus the presence of a system. Since it remains an empirical question whether communications in different dimensions co-occur, communication systems are expected to operate in various dimensions with different frequencies (cf. Simon 1973). Consequently, they may exhibit change in relation to stability.

The communication systems reproduce themselves in terms of sums of co-variations, since only through the co-variations they are able to operate. At each moment in time, *variation has to be operationalized as co-variance, and structure as the remaining variance (or auto-covariance).* The sum of the co-variance and the remaining variance of the system is conceptually equivalent to the expected information content of the system. This information content is relative to its maximum information content, and this difference is by definition equal to the redundancy. Structure, therefore, should not be equated with redundancy; it contains the remaining variance as a second dimension of the probabilistic entropy. Structure can be added to the redundancy with reference to the incoming signal, but structure can operate on (i.e., communicate with) the signal, since it contains information. If a communication system (given the redundancy) contains more structure, the *relative* weight of the communication decreases. Increase in structure leads to an increase in selection.

Selection reduces uncertainty since it normalizes the observed information with reference to the observing system. A crucial point is the negative sign introduced by the selective operation of the *a posteriori*, i.e. observing, system (cf. Brillouin 1962). If the expected information content of the sent signal was originally equal to $a$, upon selection by a structure x this value is reduced to $f(x) = a - bx$.[Note 2] (Parameter $b$ stands for a relative weight.[Note 3]) The expected information values for the incoming signals are *aligned* in a two-dimensional extension.[Note 4] In other words, selection means that the *relational* information is *positioned* by using structure as a second dimension of the information (cf. MacKay 1969; Burt 1982). Although the incoming information may be stochastically generated, its position in the system is a function of the latter's structure.

In summary, communication systems function as selective structures for the systems that communicate with them. By acting as a selector, structure packages the incoming information with reference to its own substantive uncertainty. It provides the selecting system with a (yet irreflexive) value for the incoming information by positioning it. Note that the transmitting system, and not the receiving system makes the initial selection. By receiving, a system can only select from the signals which were transmitted to it by the transmitting system ($a - bx$), and not from those sent by the sending system ($a$). The receiving system has to make a second selection, i.e., upon reception the signal $f(x)$ gets the value of $(a - bx_1)(1 - cx_2)$ or, after normalization: $f(x) = a - b'x + c'x^2$.

Each communication among communication systems implies another selection, and thereby a higher-order feedback term is added to the original signal. Selection is asymmetrical: it implies normalization by the selective device. Therefore, selection is in principle irreversible.

*Stabilization and Self-organization*

The study of the dynamics between selection and stabilization involves a second-order feedback which has to build on the cybernetics among (co-)variation and first-order selection. Stabilization is equivalent to second-order selection: which selections will be selected for stabilization?

How can one understand stabilization in the previous model? The relations of the actors can be represented as vectors, and the network as a summation of these vectors, i.e., as a two-dimensional matrix. In each cycle, the potentially co-varying relations among the actors add up to a communication system. The time dimension adds a third axis to the two-dimensional representation: matrices at different moments in time add up to a cube. If one rotates this cube ninety degrees, one can analyze structure in the time dimension, analogously to eigen-structure in the matrices at each moment in time. Let us call this reconstruction the eigen-time of the system. Eigen-time is an analytical expectation with respect to the clock of the system, just as eigen-structure is an analytical expectation with respect to structure. As is well-known, eigen-structure can be decomposed into the eigen-vectors of the system under study. Analogously, eigen-time can be decomposed into eigen-frequencies. The different frequencies of each clock can be represented as a spectrum (cf. Smolensky 1986). Note that the systems are expected to tick with different clocks.

In other words, a communication system has two selective structures if it contains information in the time dimension, and thereby organizes co-occurrences of co-variations in terms of its history. The first structure positions the information in the relations on a second dimension, and the result of this operation can be reflected on the third dimension. The two selective structures are formally equivalent, but their orthogonality implies that they are substantively completely different.

If there is no signal in a substance, there can be no reflection in a third dimension of the probabilistic entropy; the higher-order cybernetics require the lower-order cybernetics as a basis for operating. Reflection is formally a recursion of selection using a third dimension. This formal conclusion has an important theoretical implication: reflexive systems cannot be *completely* transparent to themselves, since they cannot focus on the first-order cybernetics and the second-order cybernetics at the same time and at the same place in memory. Reflexivity therefore requires internal differentiation between what is reflected, and the reflecting instance.

In a third-order cybernetics, the eigen-vectors of the matrices can additionally be combined with the eigen-frequencies of the communication system, but this requires one more degree of freedom. If the system is able to use this degree of freedom for the selection of specific relations between eigen-vectors and eigen-frequencies, it will have options to organize itself increasingly in terms of its operation, i.e., to maintain (changing) structure over time. The system is then expected to reorganize what it will consider as relevant communications and co-occurrences with hindsight, and so to self-determine its identity in the present (i.e., as a receiver). Thus, self-organization requires one more dimension than reflection.

In terms of the above spatial metaphor of a cube, one may think of a self-organizing system in terms of alternative cylinders in this cube which the system has available as internal representations of its identity. Each cylinder leads to a different expectation for the composition in the next round (cf. Dosi 1982). The operation of the self-organizing system is uncertain in a fourth dimension with reference to its three-dimensional representations. In order to maintain identity in a self-organizing system, both the communications and the co-occurrences have to be selected by the system, *and* to be reinterpreted self-referentially as information about the system. This "and" implies a third operation, and thus a fourth dimension. The subsequent operation of the system can change the relative weight of the various representations; the system must reorganize its uncertainty periodically. Note that a four-dimensional system runs by definition in a hyper-cycle that can only be observed in terms of a three-dimensional representation. If a self-organizing system no longer manages to operate in four dimensions, it will fall apart into three-dimensional representations. At a later moment, it either restores its order or vanishes.

**Table I**

Organization of concepts in relation to degrees of freedom in the probability distribution

|  | *first dimension* | *second dimension* | *third dimension* | *fourth dimension* |
| --- | --- | --- | --- | --- |
| *operation* | variation | selection | stabilization | self-organization |
| *nature or* | entropy; disturbance | extension; network | localized trajectory | identity regime |
| *character of operation* | probabilistic; uncertain | deterministic; structural | reflexive; reconstructiv | globally organized; resilient |
| *appearance* | instantaneous and volatile | spatial; multi-variate | historically contingent | hyper-cycle in space and time |
| *unit of observation* | change in terms of relations | latent positions | stabilities during history | virtual expectations |
| *type of analysis* | descriptive registration | multi-variate analysis | time-series analysis | non-linear dynamics |

*Complexity Among Systems*

Since communication systems can only operate by communicating with other communication systems, the co-variation in a communication is part of two communication systems. These systems are different in terms of what they communicate, and because of this substantive difference they can be represented as orthogonal dimensions that co-vary in the event of a communication. The relative weight of the mutual information for two communicating systems is consequently asymmetrical. In the time dimension, the asymmetry of the operation defines the past as history, and therewith the arrow of time becomes available for the reconstruction (cf. Coveney and Highfield 1990).

Reception is equivalent to stabilization of the signal, since the receiving system reconstructs the originally sent information. As noted, the receiving system selects with respect to the information available in the transmitting system; this can be modelled with the function ($a - bx + cx^2$) in which x represents the structure of the receiving system. This function can be decomposed into $(x - x_1)(x - x_2)$ if it can be provided with a value on another dimension (e.g., $f(x) = 0$). A three-dimensional system can therefore distinguish between $x_1 = a'$ as the expected information value for the signal from the sending system, and $x_2 = b'$ as the expectation of the noise introduced by the transmission.[Note 5] A two-dimensional system can hold the information, but it cannot attribute a reflexive value (i.e., a "meaning") to a signal; the signal therefore degenerates into noise. A four-dimensional system, on the other hand, is able to revise its original attribution of $a'$ to the signal and $b'$ to the noise.

Once the signal is reconstructed by the reflexive system, the original information from the sending system has become obsolete: instead of the originally sent information ($a$), the system has now the value of $a'$ available in the reconstruction, i.e., as an expectation with respect to $a$. In this process, $a$ has been completely rewritten, and in the reconstruction the noise ($b'$) has been (orthogonally) separated from the meaningful information. Without this decomposition, the signal would tend to deteriorate into noise at a next interface: a higher-order feedback term would be added ($f(x) = a - bx + cx^2 - dx^3$), and one would need a four-dimensional system for an interpretation.

Using a fourth degree of freedom a self-organizing system can reconstruct a three-dimensional signal at one further distance: in terms of (i) its expected information content, (ii) its transformation by the media through which it passed (i.e., the contextual information), and (iii) the noise. While in the case of reflexive systems, the information originally sent in the message will not reach further than two selections in the environment, a self-organizing system can reconstruct beyond a third interface. Thus, how the communication will be further processed, depends on the nature of the receiving systems.

The input can be translated into an output if the system has three degrees of freedom for the operation. But different values for $f(x)$ lead to different reconstructions; reflectors can accordingly differ in terms of the quality of the reflection. A self-organizing system can vary its aperture, and thus reflect on the quality of the various reflections. Note the paradigm change between Shannon's "fixed" communication channels and this perspective on communication systems: one can either black-box the system as a channel and consider it in terms of relations between input and output or deconstruct the same system as a reflector that uses three degrees of freedom. But if the communication channel is no longer fixed, it can be supposed to change, among other things, its reflexive function (if only by wear and tear). Therefore, it can formally be considered as self-organizing its own degeneration--i.e., it exhibits a life-cycle. Self-organization is an analytical consequence of replacing the assumption about fixed channels that can transmit with more or less noise, with communication systems that communicate information in relation to another context when disturbing the transmission.

Although the number of layers that can be reached by a communication is dependent on the complexity of the receiving system, communication systems may relate to a cascade of systems which communicate through a sequence of co-variations. Reflexive systems are consequently able to communicate among them, since they can bounce the information back and forth if they relate to the same transmitting system as the medium of communication. Self-organizing systems are additionally competent to communicate in two dimensions, since they have one more degree of freedom available for the communication. I shall argue elsewhere [Note 6] that natural language is the specific evolutionary achievement that allows us, among other things, to communicate in two dimensions (e.g., information and meaning) at the same time. Full-grown self-organizing systems are able to reconstruct the information in three dimensions (substantive information, the function of the information, and the reflexive meaning of it), but in a two-dimensional (natural) language a communicator needs a series of at least two communications (e.g., a substantive and a reflexive one) for the unambiguous transmission of information in all three dimensions.

In other words, a sending and a receiving system communicate through the transmitting system(s) which relate them. The interfaces are system/environment boundaries that are passed by the exchange of information during a communication. If one interface is involved (e.g., between the sending and transmitting systems or between the transmitting and receiving systems) the signal is selected, and the communication systems have to be structurally coupled; in the second interface, the signal is transmitted, and the communicating systems are operationally coupled, i.e., by the operation of the shared structurally coupled one; in the third interface, the signal can be translated, and the systems are loosely coupled, since the reflexive system is able to select a specific signal as an input among the various signals arriving from operationally coupled systems. Self-organizing systems can reach beyond this third interface, and consequently they are able to revise their reconstruction with respect to the systems which have been selected as operationally or loosely coupled. Consequently, they are able to revise what will be considered as noise and what as signal, and therefore they are able to learn. As noted, all interfacing is asymmetrical, and the *a posteriori* perspective of the reception determines the time axis.

One should understand these statements at the generalized level, i.e., regardless of the specification of what the researcher will consider as a communication system, a subsystem, or a supersystem. For example, operational coupling can either be considered as interaction among the subsystems of one system to which they both structurally belong, or as interaction among two systems that are specified as structurally coupled to a communication (super-)system between them. Operationally coupled systems cannot communicate without using a system to which both of them are structurally coupled.

Note that our common-sense notion of communication between a sender and a receiver implies two communications in the mathematical theory of communication: one between the sending system and the transmitting system and one between the latter and the receiving system. Through the first layer of structurally coupled communication systems, each system can communicate with all the systems which structurally belong to the same network. Each communication of a communicating agent with the network can therefore affect all the operationally coupled receiving systems. A receiving system can only redirect the information if it is able to decompose the two dimensions of the information contained in the message in terms of an expectation with respect to the signal and the noise. Otherwise, the signal tends to perish into noise. A self-organizing system can additionally interpret information from communication systems to which it is itself only loosely coupled (e.g., through an organization).

For example, if a human psyche wishes to move, it signals through the nervous system to a muscle to contract, but the muscle as a specific structure already contains the information of how to perform this function in coordination with other bodily functions. The psyche cannot intervene directly in the interface between the muscle and the bone, unless the muscle allows for this coupling. Deeper layers of communication within the body, like the delivery of nutrients for the contraction, are increasingly beyond control. Various communication systems communicate in ever deeper layers of communication. The number of systems in the cascade of communications is an evolutionary variable (Maturana and Varela 1988). As noted, stabilization requires at least three dimensions, and therefore reflexivity. Reflection continuously filters the noise, and therefore evolutionary achievements can be maintained as nearly decomposable systems (Simon 1969).

Communication systems which cannot communicate through shared environments, because they are more than two layers away, may sometimes be made to communicate through again completely different communication systems. For example, although as human beings we are usually unable to communicate internally with the various circulations of matter through our body, we can directly inject a metabolite into the blood with a hypodermic syringe. A self-organizing system can intervene by redefining which systems will be operationally coupled and which only loosely, since it has a fourth degree of freedom available for making an internal representation of three-dimensional (input/output) systems.

*Complexity in the Time Dimension*

In the previous section, I focussed on the evolutionary complex as an architecture of nearly decomposable systems, and in this section I shall discuss its development over time. As noted, the two axes are formally equivalent, since both function as selectors, but selections in either dimension have given rise to a different semantics. Communication in the time dimensions has to be considered, for example, not in terms of the co-variance, but as a co-occurrence.

If one represents the communicating systems as the row vectors of a matrix, communication finds its origin in the co-occurrence of these vectors along the column dimension. Let us call this dimension $i$. If originally the rows are just stacked upon one another, the result is a segmented communication system. If the communications are ranked in the vertical dimension, one gets a stratified communication system, and grouping of the rows leads to a differentiated communication system. Grouping presupposes that the ranking is maintained over time; ranking presupposes that the stacking is maintained over time. Ranking, thus, is a special (simple) case of grouping. Grouping implies the addition of another dimension, i.e., a grouping variable $j$. In the case of stratification, the grouping variable only counts the rank. For example, in a stratified social system a person is allowed to say something if it is his or her turn.

Co-occurrence, however, is an idealization by a receiving system, since all events occur discretely if looked upon with sufficient precision. A system can consider the different occurrences as co-occurrences only reflexively. In a stratified system, the communications are ranked at a reflexive centre, but not yet grouped. Reflection on the distinction between this centre and the periphery using a fourth dimension induces differentiation. In a self-organizing system, this differentiation is reflexively adjusted to functionality for the system. In summary: segmentation requires co-variation in two dimensions; stratification requires stabilization in three dimensions, and consequently a difference between the reflecting instance and the reflected substance; functional differentiation requires specificity of the reflection with reference to the organization of the system, and therefore four dimensions.

A system can reflect in the third dimension the traces that the individual occurrences (first dimension) leave in a (two-dimensional) substance. As soon as the communications leave traces, one has to assume substance (and therefore a two-dimensional extension) of a communication system in which the communications can leave traces. The traces enable a three-dimensional system for reflexively distinguishing co-occurrences in an interval with a certain band-width. From this perspective structure can be considered as the sum of all the communications in the reconstructed interval on the time dimension. As soon as the emerging communication system contains structure, this structure becomes significant for the further development of the system at the next moment in time, since structure determines selection (see above).

Thus, a substance that is left with options for change will exhibit a tendency to develop further also in the third dimension (cf. Arthur 1988). Analogously, a three-dimensional system will tend to become four-dimensional, and thus to exhibit a life-cycle. The continuous dissipation of probabilistic energy drives all entropical systems eventually towards their deaths. Accordingly, a system's capacity to transmit will always tend to degenerate into a capacity to transform. If one wishes to stabilize a channel for the transmission, an engineer has to *fix* it for this purpose.

The higher-order system is more stable (i.e., better buffered because it contains more reflexive filters) than the lower-level one, and therefore it tends "to take over." This is not Darwinian "survival of the fittest," since the lower-level system is encompassed in the higher-level one. For example, when grouping prevails, there may be parts of the system that are not yet grouped, but segmentation has in this stage disappeared. This growth pattern is well-known from biology: once the *morula* grows so large that there is a need for synchronization among cells no longer directly adjacent, order emerges. The event of a cell-cleavage is asymmetrically communicated to neighbouring cells, and triggers there a further cleavage. At first, this order is only rank-order, i.e., stratification or, in biological terms, "polarization" (*gastrula*). But the next stage (*blastula*) can be defined as the phase after which undifferentiated cells cease to occur.

As long as the windowing of communication systems on each other remains direct, there is no evolutionary order. The sequencing within the system induces order. If the repetition over time leaves traces, some traces will grow more dense than others. Thus, the groupings can begin to differentiate. If the system is to organize itself, it must build on differentiations that are functional for its further development. Functional differentiations become "locked in," i.e., pathways emerge which are preferentially used for further traces. Thus, the system (once materialized in a substance which can retain traces) has a tendency to develop a history, and thus to become potentially reflexive and differentiated.

If a fourth degree of freedom can be made available (e.g., by relating to another context) specific resonances between the first-order and the second-order cybernetics become possible. If one of these resonances happens through stochastic variation, the functionality of the differentiation spreads, since the remainder of the system consists of groupings that are not yet functional. As noted, in functionally differentiated organisms, undifferentiated cells cease to occur.

Again, but in a more formalized terminology: if the non-differentiated medium of communication is indicated with *i* (see above), the differentiated medium must be indicated with *ij*, since a grouping variable *j* is added. Once there is differentiation, the value for *j* may not yet have been specified for some (yet undifferentiated) communications. Additionally, if the one functionally differentiated subsystem, which for example communicates in *i* and *j*, communicates with another functionally differentiated subsystem (in *i* and *k*) of the same system, this does not imply de-differentiation and thus communication in only *i*, but the emergence of communication among *i*, *j*, and *k*. De-differentiation can occur only *locally*, when *j* and *k* cancel one another like in patterns of interference. In a chaology (in contrast to a cosmology) integration means an increase in complexity, and only local specimens can be found that are not yet differentiated, and therefore can carry the next generation. For existing systems, there is no return, since the more complex system is *a posteriori* to the less complex one.

Let me use the example from the previous section. First, if a human being wishes to move, it needs an interface which not only makes the organs involved (nerves, muscles, bones, etc.) recognize one another as tissue of the same animal (*i*), but which also structures the communication between, for example, the nervous system (*ij*) and the motoric apparatus (*ik*). This operational coupling requires a structural coupling into an interfacing system (e.g., a synapse) which "knows" how to translate input into output; by structurally doing so, the interfacing system composes a three-dimensional system (*i*, *j*, and *k*). Only a three-dimensional system can contain sufficient complexity to perform translations between differentiated subsystems. Additionally, this three-dimensional system exists over time, and thus performs a life-cycle (with the supersystem *i*...).[Note 7]

Note that the functionally differentiated system can generate more complexity at its internal interfaces. Whether these new complexities can also be stabilized, depends on the extent to which they tend to be repeated, and whether they can be locked into a relevant context. Once a threshold is passed, the traces can be inscribed into the higher-order system as a dimension.

*Summary and Conclusions*

A system can deconstruct a signal of one-lower dimensionality than it has available, since it needs the additional dimension in order to provide the signal with a value. In the second dimension we have called this the positioning of the relational information; in the third dimension reflection; and in the fourth dimension it has occasionally been called reflexivity, but I have suggested that it should be clearly distinguished from reflection by calling it self-organization. The underlying operation among the various dimensions, however, is identical: the incoming information is always mutual information with reference to the total information content of a system. Normalization can only be achieved if the receiving system is able to compute this fraction as a percentage co-variance (cf. Leydesdorff 1992). This presumes an internal representation, and hence the projection onto another dimension.

A three-dimensional system can reconstruct input into output; a four-dimensional system can reconstruct in three dimensions. For example, a bird can build a nest in three dimensions. An electronic amplifier is also a three-dimensional system, but it is based on cultural evolution--i.e., a reconstruction at the level of the social system--and this further complicates the issue. However, the capacity for ordering the traces of occurrences into co-occurrences and sequences of events is a capacity of the medium of communication. Both the biological cell and the properly constructed electronic circuit are able to retain information, while for example our spoken communications in the air are volatile (if not registered otherwise).

Whether higher-order systems with the capacity to retain information emerge, and can further develop evolutionarily depends on whether sufficient probabilistic entropy is generated in each relevant context (Swenson 1989). If probabilistic entropy is generated in one context, the distribution of this noise can become increasingly skewed, if only as a result of stochastic drift (Arthur 1988). If pronounced enough, this information may be recognized as a signal by another system. As soon as the other system can process this signal, a co-evolution can begin, and a higher-order stability can be generated. Subsequently, this communication will be locked into the higher-order system, since the higher-order system is more stable than the lower-level one. Of course, the higher-order system can again decay (sometimes rapidly), but when this will occur is determined at this level of control, and only probabilistically influenced by activities at lower levels.

*The Possibility of Artificial Evolution (Discussion)*

The capacity of a system for ordering eigen-values and eigen-frequencies into a third-order cybernetics requires the summation of previous operations, and thus control over a period of time, i.e., memory has then to be available as a property of the medium. Only systems with this function can decide internally how to attribute communications to different parts of the communication system. A computer system has this function in addition to input and output channels. But the difference between living systems and (traditional) computer systems can now be specified in terms of whether the fourth dimension is available as a degree of freedom. Living systems can reprogram *a posteriori* in response to experiences, while computer systems with a Von Neumann architecture have to be programmed (and time stamped!) *ex ante*, however much ingenuity may have been built in for making them "learn" from later experiences. Thus, the difference is not the reflexivity, but whether the reflexivity of the system has to be specified *ex ante* or can be varied *ex post*. The living system takes its current situation as a point of reference, and may make irreversible transitions on the basis of incoming information which were highly unlikely *a priori*.

Programming all possible transitions in the higher-order dimension *ex ante* makes the programming task non-polynominal complete, and therefore uncomputable (cf. Penrose 1989). But can't we build computers with higher dimensionalities? In a Von Neumann architecture, input is essentially translated into output. Thus, the architecture of the system remains three-dimensional. The specification of an additional do-while loop in the software enables us to simulate higher-order complexities, but not to map and to reconstruct them. In parallel and distributed processing (PDP), however, one adds an extra dimension to the hardware of the system, and therefore software can be written so that this four-dimensional system self-organizes three-dimensional reflections, e.g., in a Boltzmann-machine that can recognize patterns (cf. Hinton & Sejnowski 1986).

The problem of the reconstruction of the living (cf. Langton 1989 and 1992) would need five dimensions, since reconstructions require one more dimension than contained in the subject under study. A four-dimensional system is able to reconstruct a three-dimensional system to the extent that it can design one, but it can only reconstruct a four-dimensional system to the extent that it can develop and improve its mental mappings of it, given the specification of a perspective. A five-dimensional system would be able to reconstruct a four-dimensional one. Thus, the project of "artificial life," "artificial intelligence," and "artificial evolution" (Anderson 1992) by using computers is tractable, but the problem is currently uncomputable given the availability of hard- and software.

[(return)](#)

**Notes**

1. (Mutual) information content and (co-)variance are differently defined, but they are both measures of the communicated uncertainty (cf. Theil 1972). Although not equal, the two concepts are equivalent for the discursive argument.

2. For didactic purposes, I use the idealization of the continuous functions here, although I shall argue below that all information is discrete if inspected with sufficient precision.

3. Since each communication system is able to update only in terms of of its own substance, a normalization is required.

4. If $f(x) < 0$, the signal vanishes upon incorporation into the transmitting system (e.g., by absorption).

5. The orthogonality among the dimensions provides us with the other equation required for this decomposition.

6. L. Leydesdorff, "The Sociological and the Mathematical Theory of Communication," (in preparation).

7. If a muscle is denervated, control by the higher-level system is released, and the sensitivity of the lower-level system for disturbances is generally increased.

[(return)](#)

**References**

Anderson, E. S. (1992). *Artifical Economic Evolution and Schumpeter*. Aalborg: Institute for Production, University of Aalborg.

Arthur, W. B. (1988). Competing Technologies, in: Dosi, G., *et al.* (eds.), *Technical Change and Economic Theory*. London: Pinter, 590-607.

Brillouin, L. (1962). *Science and Information Theory*. New York: Academic Press.

Burt, R. S. (1982). *Toward a Structural Theory of Action*. New York, etc.: Academic Press.

Coveney, P., and R. Highfield (1990). *The Arrow of Time*. London: Allen.

Dosi, G. (1982). Technological paradigms and technological trajectories, *Research Policy*, **11**, 147-62.

Hinton, G. E. and T. Sejnowski (1986). Learning and Relearning in Boltzmann Machines, in: Rumelhart, D. E., J. L. McClelland, and the PDP Group, *Parallel Distributed Processing*. Cambridge, MA/London: MIT Press, Vol. I, pp. 282-317.

Langton, C. G. (1989). *Artificial Life*, Redwood City, CA: Addison Wesley.

Langton, C. G., C. Taylor, J. Doyne Farmer, and S. Rasmussen (1992). *Artificial Life II*. Redwood City, CA: Addison Wesley.

Leydesdorff, L. (1992). Knowledge Representations, Bayesian Inferences, and Empirical Science Studies, *Social Science Information*, **31**, 213-37.

Maturana, H. R., and F. J. Varela (1988). *The Tree of Knowledge*. Boston: New Science Library.

Penrose, R. (1989). *The Emperor's New Mind*. Oxford: Oxford University Press.

Shannon, C. E. (1948). A Mathematical Theory of Communication, *Bell System Technical Journal*, **27**, 379-423 and 623-56.

Simon, H. A. (1969). *The Sciences of the Artificial*. Cambridge, MA/ London: MIT Press.

Simon, H. A. (1973). The Organization of Complex Systems, in: H. H. Pattee (ed.), *Hierarchy Theory. The Challenge of Complex Systems*. New York: George Braziller, pp. 1-27.

Smolensky, P. (1986). Information Processing in Dynamical Systems: Foundation of Harmony Theory, in: D. E. Rumelhart, J. L. McClelland, and the PDP Group, *Parallel Distributed Processing*. Cambridge, MA/London: MIT Press, Vol. I, pp. 194-281.

Swenson, R. (1989). Emergent Attractors and the Law of Maximum Entropy Production: Foundations to a Theory of General Evolution, *Systems Research*, **6**, 187-97.

→